\documentclass[aps,twocolumn]{revtex4}
\usepackage{graphicx}

\newcommand{\be}{\begin{equation}}
\newcommand{\ee}{\end{equation}}
\newcommand{\bea}{\begin{eqnarray}}
\newcommand{\eea}{\end{eqnarray}}

\begin{document}

\title{First passage times and asymmetry of DNA translocation}
\author{Rhonald C. Lua, Alexander Y. Grosberg}
\affiliation{Department of Physics, University of Minnesota \\
116 Church Street SE, Minneapolis, MN 55455}

\begin{abstract} Motivated by experiments in which single-stranded
DNA with a short hairpin loop at one end undergoes unforced
diffusion through a narrow pore, we study the first passage times
for a particle, executing one-dimensional brownian motion in an
asymmetric sawtooth potential, to exit one of the boundaries. We
consider the first passage times for the case of classical
diffusion, characterized by a mean-square displacement of the form
$\left< (\Delta x)^2\right> \sim t$, and for the case of anomalous
diffusion or subdiffusion, characterized by a mean-square
displacement of the form $\left< (\Delta x)^2\right> \sim
t^{\gamma}$ with $0<\gamma<1$. In the context of classical
diffusion, we obtain an expression for the mean first passage time
and show that this quantity changes when the direction of the
sawtooth is reversed or, equivalently, when the reflecting and
absorbing boundaries are exchanged. We discuss at which numbers of
`teeth' $N$ (or number of DNA nucleotides) and at which heights of
the sawtooth potential this difference becomes significant. For
large $N$, it is well known that the mean first passage time
scales as $N^2$. In the context of subdiffusion, the mean first
passage time does not exist. Therefore we obtain instead the
distribution of first passage times in the limit of long times. We
show that the prefactor in the power relation for this
distribution is simply the expression for the mean first passage
time in classical diffusion. We also describe a hypothetical
experiment to calculate the average of the first passage times for
a fraction of passage events that each end within some time
$t^{\ast}$. We show that this average first passage time scales as
$N^{2/\gamma}$ in subdiffusion.
\end{abstract}

\maketitle

\section{Introduction}

Recent studies on the passage of DNA through narrow channels,
apart from being very interesting in their own right, have been
motivated by the exciting possibility of developing a practical
technique to characterize and sequence DNA.  These single molecule
experiments were pioneered by Kasianowicz, Brandin, Branton and
Deamer \cite{Kasianowicz} using a protein called
$\alpha$-hemolysin as a pore embedded in a lipid membrane.  When
voltage is applied across the membrane, ion current can be
detected. When the DNA chain is inside and blocking the channel,
the current is suppressed. By measuring the blockage time one can
gather information about the voltage driven dynamics of DNA
translocation.  A large amount of exciting data have been
accumulated through such experiments
\cite{Akeson,Meller1,Meller2,Meller3}.  Recently, another group of
experiments was performed using solid state nano-pores for
voltage-driven DNA translocation \cite{Golovchenko,Storm}.

\begin{figure}
\centerline{\scalebox{0.5}{\includegraphics{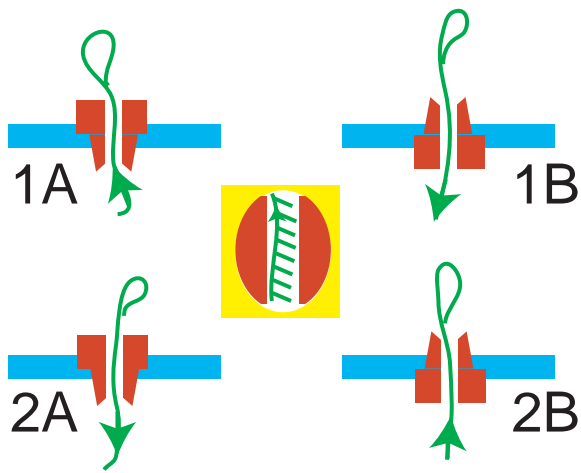}}}
\caption{(Color online) The four possible relative orientations of
DNA (``key'') and pore (``keyhole'').  In a similar figure from
Lubensky and Nelson (reference \cite{Lubensky}, figure 7), the
single-stranded DNA has no loop. Instead, the four cases were due
to the various relative orientations of DNA, pore and an applied
electric field.  Arrows in our figure show the direction from the
$5^{\prime}$ to the $3^{\prime}$ end in the DNA.  Inset in the
middle shows schematically the tilted bases. The analysis in our
work compares the passage times for case 1A with case 1B (or 2A with
2B) in which the relative orientation between the DNA bases and
pore is identical. In contrast, the experiments in
\cite{Meller_explanation} study cases 1A and 2A, where the DNA
enters the pore from the same ($cis$) side. }
\label{fig:dnaporecases}
\end{figure}

On the theoretical side, much effort was placed in understanding
the entropic barrier associated with translocation of a long
polymer \cite{Park,Muthukumar}.  The electrostatic barrier
associated with charged DNA penetration through the low dielectric
constant membrane has been recently discussed theoretically
\cite{ZhangKamenevLarkinShklovskii}.  Another interesting question
is about friction: whether friction on the small piece of DNA
passing through the narrow channel is larger or smaller than the
friction experienced by the large DNA coils outside the membrane.
To this end, Lubensky and Nelson \cite{Lubensky}, assuming the
channel friction dominance, were able to account for the observed
bimodal distribution of the passage times for single-stranded DNA.
In their model, each base preferentially tilts towards one end of
the single stranded DNA chain ($5^{\prime}$ or $3^{\prime}$. See
figure \ref{fig:dnaporecases}.). Each peak in the distribution of
passage times therefore corresponds to a particular end entering
the channel first, the channel being just wide enough for a single
strand to pass through.  More recently, Kantor and Kardar
\cite{KantorKardar} and then Storm et al \cite{Storm} argued that
for sufficiently long DNA there must be a crossover to the regime
of domination of the out-of-channel friction, in which case
translocation should become subdiffusive in character (with
displacement growing slower than $t^{1/2}$ with time $t$).

A new spin is added to the story by the experiment by Meller and
coworkers \cite{Meller_explanation}.  These authors devised an
experiment involving DNA having a string of identical bases
(adenine) in the single-stranded portion and a hairpin loop at one
end (figure \ref{fig:dnaporecases}) held in place by bonding of
complementary bases. Like double-stranded DNA, the hairpin cannot
enter the channel (a transmembrane pore of $\alpha$-Hemolysin).
The hairpin therefore constrains the DNA to enter the pore with
its single-stranded end, as well as preventing the entire DNA from
crossing the membrane.  In their experiment, the DNA, driven by an
applied voltage, enters the pore with its single-stranded end.
Thereafter, once current is blocked by the DNA, the voltage is
either switched off, in which case the DNA diffuses freely
(non-driven), or the sign of the voltage is flipped, in which case
the DNA is pushed back.  Moreover, by making two DNA samples, with
the hairpin loop at opposite ends, it is possible to observe DNA
sliding away from the pore in two opposite directions along the
DNA contour, and the observation suggests that DNA escapes in one
direction faster than in the other.

In the experiment \cite{Meller_explanation}, by measuring the
so-called ``survival probability'' $S(t)$, which is the
probability that a DNA molecule will stay in the pore as a
function of the waiting time, it was determined that the
voltage-free dynamics of the $3^{\prime}$ threaded molecules is
about two times slower than the corresponding diffusion of
$5^{\prime}$ threaded molecules having the same sequence.
Importantly, in both cases the DNA was threaded from the same side
of the pore (called the $cis$-side of $\alpha$-HL).  To delineate
the underlying mechanism responsible for the observed dynamics,
the authors of the work \cite{Meller_explanation} performed
all-atom molecular dynamics simulations, which independently
confirmed the experimental results for driven DNA. The simulations
also showed that the confinement of the DNA bases in the
$\alpha$-HL pore results in an even stronger (compared to a free
DNA) tilt of the bases with respect to the DNA backbone towards
the $5^{\prime}$ end.

Authors of the work \cite{Meller_explanation} phenomenologically
interpret their data by assigning two different diffusion
constants for the two separate experiments in which the same DNA
is placed in the channel in two possible orientations. This
interpretation is justified by the fact that the interactions
between the DNA bases and the pore are different in these two
cases (perhaps via different barrier heights within the framework
of a sawtooth potential landscape discussed below).

\begin{figure}
\centerline{\scalebox{0.5}{\includegraphics{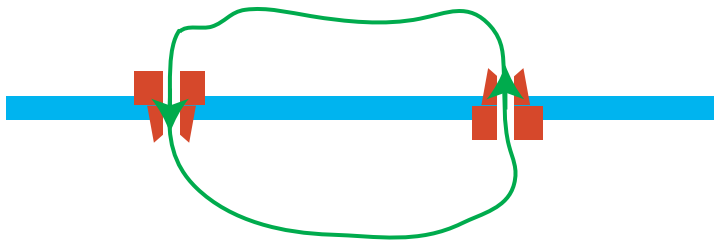}}}
\caption{ (Color online) This arrangement of DNA and pores would
have acted as a \textit{perpetuum mobile} if the stationary
diffusion coefficient was asymmetric. This shows that it cannot be
asymmetric, the symmetry being a requirement of thermodynamics.}
\label{fig:perpetuum_mobile}
\end{figure}

There is a temptation to summarize the experimental findings of the
work \cite{Meller_explanation} in one sentence (although no one
made this mistake, including \cite{Meller_explanation}): DNA
diffuses in one direction faster than in the other. Indeed, the
observed asymmetry of dynamics is consistent with the tilt of the
nucleotides with respect to the main DNA chain. This asymmetry
then seems easy to understand if the analogy is made with petting
a cat along or against the grain of its fur; the cat responds very
differently in the two cases (presumably because it experiences
very different friction). Another, possibly even more obvious,
analogy would be carrying a Christmas tree top first or base first
through a narrow door; one again encounters very different
resistance in the two cases. The point is that such analogies and
interpretations are only possible for the driven DNA motion,
particularly for the system far into the non-linear regime (in
terms of force-velocity relation), whereas for the portion of
experiments in the work \cite{Meller_explanation} involving freely
diffusing DNA such analogies and interpretations would be
\emph{wrong}; it is not surprising then that the authors of the
work \cite{Meller_explanation} did not use such analogies and
interpretation for the freely diffusing DNA.  Indeed, for free
diffusion, the friction coefficient (averaged over the scale well
exceeding a single base) moving in one direction and in the opposite
direction must be the same, as follows from the Onsager symmetry
relation, and the assumption of asymmetric friction would be a
grave mistake. Although no one actually made this mistake,
including \cite{Meller_explanation}, it is worth emphasizing why
an assumption of asymmetric friction would be a mistake. Indeed,
if we only imagine that DNA (not driven by any applied voltage!)
diffuses in one direction faster than in the other, then we can
easily build a \textit{perpetuum mobile} (see figure
\ref{fig:perpetuum_mobile}) moving indefinitely long through time
at the expense of thermal energy from the thermal bath, which is,
of course, impossible.  In other words, freely diffusing DNA, when
it is already in the pore, in contrast to a (heavily driven!)
Christmas tree through a door, must have the same friction
coefficients when the DNA moves in either direction.

What is nice is that the experimental findings and their
interpretation in the work \cite{Meller_explanation} are in fact
in perfect agreement with this thermodynamic analysis. In order to
make this reconciliation very clear, we immediately refer to the
symmetry analysis in figure \ref{fig:dnaporecases}.  Notice that
the pore itself is asymmetric (its crystallographic structure is
known \cite{structure}), the DNA backbone is also asymmetric (from
$3^{\prime}$ end to $5^{\prime}$ end), and the loopy end creates
further asymmetry.  This gives four possible orientations of the
pore and the DNA with the loop: two possibilities arise from two
different mutual orientations of the DNA backbone with respect to
the pore (indicated by the numbers 1 and 2 in figure
\ref{fig:dnaporecases}), and for each of these two orientations
there are two possibilities to place the blocking loop (indicated
by the letters A and B in figure \ref{fig:dnaporecases}). This
symmetry analysis, as shown in figure \ref{fig:dnaporecases}, is
reminiscent of the symmetry analysis in the paper \cite{Lubensky},
except we have no electric field, but instead have loops at the
DNA ends.

We can now say that in any one of the arrangements, from 1A, 1B,
2A or 2B, the DNA must experience the same friction moving up or
down the pore; friction going up equals friction going down. At
the same time, the friction in configurations 1A or 1B can be
different from friction in configurations 2A or 2B, and they are
likely to be different. That is why the work
\cite{Meller_explanation} assigns two different diffusion
constants to the two DNA-pore mutual configurations (1A and 2A).
By contrast, the loop itself likely has no effect on the friction
or diffusion coefficient, so we expect that the diffusion
coefficient should be the same for configurations 1A and 1B (same
goes for 2A and 2B).  In other words, there should be two distinct
diffusion coefficients, not four.  We shall argue in this work
that, nevertheless, there will be four different diffusion times
corresponding to the four configurations in figure
\ref{fig:dnaporecases}.

To explain our approach, it is convenient to adopt a terminology
in which, instead of considering diffusion of the DNA chain, we
consider diffusion of the passage point along the DNA contour.
Following Lubensky and Nelson \cite{Lubensky}, we consider a
simple model in which asymmetry is presented in the underlying
potential landscape.  For simplicity, we model it with a sawtooth
profile.
The two orientations of the asymmetric potential (relative to the
boundary conditions) correspond to the two possible placements of
the blocking loop for a given orientation between DNA and pore
(e.g. 1A and 1B in figure \ref{fig:dnaporecases}).

Like Lubensky and Nelson \cite{Lubensky}, we focus on the first
passage time, which is the time it takes for the initially
fully-`plugged' DNA to completely `unplug' from the pore.  In
other words, it is the time needed for the diffusing particle (or
a random walker) to arrive for the first time on the open end of
the DNA, or to one end of the $(0,L)$ interval, provided that a
reflecting boundary condition is imposed at the opposite end.

We would like to emphasize the fundamental difference between
asymmetric diffusion, which is prohibited by thermodynamics, and
symmetric diffusion over the asymmetric potential landscape.  It
is well known, and we show it explicitly in appendix
\ref{sec:steadystate}, that stationary diffusion remains symmetric
despite the asymmetry of the underlying potential landscape, thus
making nonfunctional the \textit{perpetuum mobile} design of
figure \ref{fig:perpetuum_mobile}.

In this paper, we compute the mean first passage times (MFPT)
corresponding to cases 1A and 1B in figure \ref{fig:dnaporecases}
(or to cases 2A and 2B). We consider the brownian motion of a
particle diffusing classically in an asymmetric sawtooth potential
$U(x)$ and in the inverted or reversed version of the potential
(figure \ref{fig:sawtooth}).
This model neglects the entropic barrier (of order $k_B T \ln N$) presented by
the DNA coils on both sides of the pore \cite{Park,Muthukumar}, but through
the consideration of subdiffusion it does take into account the extra
friction created by those coils \cite{KantorKardar1}.
From the results, we discuss when the
difference between the two times is significant. (Note that since
we know very little about the details of the interactions between
the DNA bases and the pore, we cannot determine if case A in
figure \ref{fig:sawtooth} corresponds to case 1A in figure
\ref{fig:dnaporecases}, and case B in figure \ref{fig:sawtooth}
corresponds to case 1B in figure \ref{fig:dnaporecases}, or if it
is the other way around.)

Since DNA translocation is ultimately not classical diffusion, but
rather subdiffusion \cite{Storm,KantorKardar}, we consider also
the first passage times for the subdiffusion in the presence of an
asymmetric potential.  In general, the first passage time for
subdiffusion was recently a matter of considerable interest and
dispute in the literature \cite{Gitterman1,Yuste,Gitterman2}.  It
is now understood \cite{WhenAnomalous,Restaurant, PhysicsWorld}
that the \textit{mean} first passage time diverges for
subdiffusion, because a subdiffusing walker tends to remain too
long on the place that it once reached.  Accordingly, we look at
the probability distribution for the first passage times (DFPT),
and concentrate on its tail at long times.  We found that this
tail is very different for the two potentials, and the difference
turns out to be expressed through corresponding mean first passage
times for classical diffusion. With this knowledge, we construct
an average first passage time from a subset of passage events and
show that this average scales as $N^{2/\gamma}$. The result also
exhibits the asymmetry between cases 1A and 1B (or 2A and 2B) just
as in the case of classical diffusion.

From our discussion we make the prediction that the observed
passage times for the four possible mutual orientations of the
pore and the DNA will all be different.

\section{Results}

\subsection{Classical diffusion}
For treating classical or normal diffusion, one often starts with
the Fokker-Planck (FPE) equation (in this context also frequently
called Smoluchowsky equation)
\begin{equation}
\frac{\partial P(x,t)}{\partial t}=D\frac{\partial}{\partial x}
e^{-U(x)} \frac{\partial}{\partial x} e^{U(x)} P(x,t) \label{FPE}
\end{equation}
giving the time-evolution of the probability density $P(x,t)$.
Here $D$ is the usual diffusion constant and we have set $k_BT=1$.
The FPE yields the Boltzmann distribution for $P$ in the
steady-state, as well as giving the linear relation between the
mean-squared displacement and time in the absence of external
forces.

\begin{figure}
\centerline{\scalebox{0.5}{\includegraphics{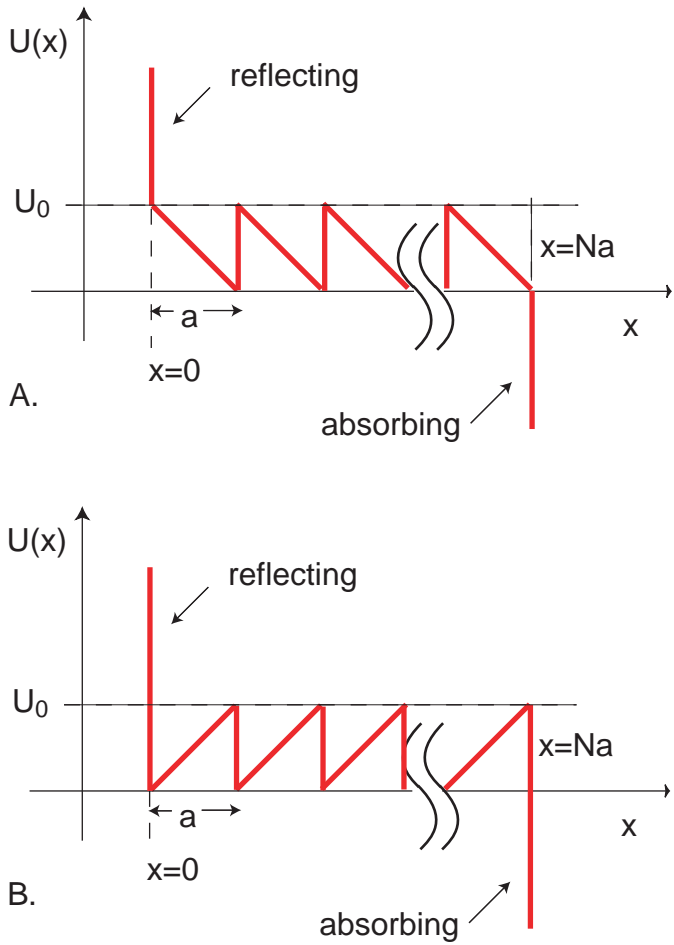}}}
\caption{(Color online) The sawtooth potential of $N$ teeth with a
reflecting boundary at $x=0$ and an absorbing boundary at
$x=L=Na$, illustrated for two different directions of asymmetry in
the sawtooth. The reflecting boundary corresponds to the inability
of the DNA hairpin to pass through the pore.} \label{fig:sawtooth}
\end{figure}

In calculations involving the first passage time, it would be
convenient to consider the equivalent problem of first passage to
either $x=L=Na$ or $x=-L=-Na$, where the potential $U(x)$ for
$x>0$ is as illustrated in figure \ref{fig:sawtooth}, while the
potential for $x<0$ is $U(x)$ for positive $x$ reflected about the
vertical axis. With this picture, the probability for the particle
to still be `alive' at time $t$, also called the survival
probability $S(t)$ (and measured in experiment
\cite{Meller_explanation}), is given by
$S(t)=\int_{-L}^{L}P(x,t)dx$. The distribution of first passage
times $F(t)$ is calculated from $S(t)$ via $F(t)=-\frac{\partial
S(t)}{\partial t}$. This gives the following expression for the
mean first passage time $\tau(x_0)$ \cite{Redner}:
\begin{eqnarray}
\tau(x_0)&=&\int_0^{\infty}tF(t)dt\nonumber\\
&=&\int_0^{\infty}S(t)dt\nonumber\\
&=&\int_0^{\infty}\int_{-L}^{L}P(x,t)dxdt \label{MFPT}
\end{eqnarray}
where $x_0$ is the initial position of the particle,
$P(x,0)=\delta(x-x_0)$.

It can be shown that $\tau(x_0)$ satisfies an ordinary
differential equation \cite{Pontryagin,FiftyYearsKramers,Risken}
(derived in appendix \ref{sec:equation_for_tau}). The solution of
this differential equation for a sawtooth potential $U(x)$ is
outlined in appendix \ref{sec:solution}. For the particle
initially located at the origin ($x_0=0$), the mean first passage time to
reach $x=L=Na$ is given by,
\begin{equation}
\tau_A = \alpha \frac{L^2}{2D} - \beta \frac{aL}{D}
\label{tauclassic0}
\end{equation}
for the potential in figure \ref{fig:sawtooth}A, and
\begin{equation}
\tau_B = \alpha \frac{L^2}{2D} + \beta \frac{aL}{D}
\label{tauclassic1}
\end{equation}
for the potential in figure \ref{fig:sawtooth}B.
Here, we have defined the coefficients
\begin{eqnarray}
\alpha &=& \left(\frac{\sinh{(U_0/2)}}{U_0/2}\right)^2 \\
\beta &=& \frac{\sinh{(U_0)}-U_0}{U_0^2} .
\end{eqnarray}
Expression (\ref{tauclassic1})
can be obtained from (\ref{tauclassic0}) by flipping the
sign of $U_0$.

From the results (\ref{tauclassic0}) and (\ref{tauclassic1}), it
is clear that $\tau_A < \tau_B$.  Physically, the inequality
$\tau_A<\tau_B$ may be obvious for the case $N=1$ in figure
\ref{fig:sawtooth}, in which a particle has to surmount a single
barrier in order to get to $x=L$ in case B, while there is no
barrier in case A. In general, for a given $N$, the potential in
figure \ref{fig:sawtooth}A involves $N-1$ barriers, while the
potential in figure \ref{fig:sawtooth}B involves $N$ barriers. In
fact, it is easy to show that in the limit $U_0 \gg 1$ ($U_0/k_BT
\gg 1$ in more conventional units) we have $\tau_A(N+1) \simeq
\tau_B(N)$, where the arguments indicate the number of teeth in the
sawtooth potentials.

For the long DNA, when $L \gg a$ or $N \gg 1$, the leading terms
in both $\tau_A$ (\ref{tauclassic0}) and $\tau_B$
(\ref{tauclassic1}) are proportional to $N^2$, as one would expect
for diffusion times.  To this leading order, first passage times
$\tau_A$ and $\tau_B$ obey the symmetry in diffusion
and are the same.  It is
in the subleading terms (proportional to $L$) that the two times
differ.  Let us stress that the difference between $\tau_A$ and
$\tau_B$, which is of order of $1/N$ in a relative sense, is
entirely due to the boundary conditions and the situation at the
ends of the diffusion region.

\subsection{Anomalous diffusion}

Anomalous diffusion is characterized by the occurrence of a mean
square displacement of the form $\left< (\Delta x)^2\right> \sim
t^{\gamma}$, where $0< \gamma < 1$ in subdiffusion; traditionally
\cite{Metzler}, this is written in the form
\begin{equation}
\left< (\Delta
x)^2\right>=\frac{2D_{\gamma}}{\Gamma(1+\gamma)}t^{\gamma}
\end{equation}
where $D_{\gamma}$ is a generalized diffusion constant and
$\Gamma(x)$ is the gamma function. For $\gamma=1$ one recovers the
usual result for classical diffusion. It can be shown
\cite{Metzler} that this form for the mean square displacement can
be obtained from a generalized version of equation (\ref{FPE})
called the fractional Fokker-Planck equation (FFPE).  This
equation is described in appendix \ref{sec:fractional_FFPE}.

Although up to this point we have ignored the interactions of the
DNA bases outside the pore, it seems reasonable to speculate that
their effect is to slow down the translocation. Thus one might be
able take these interactions into account phenomenologically by
positing a value of $\gamma$ corresponding to the subdiffusive
domain $0 < \gamma <1$. (Reference \cite{WhenAnomalous} lists
possible sources of waiting time distributions leading to
anomalous diffusion).

It is shown below and in references
\cite{Yuste,Gitterman2,WhenAnomalous,Restaurant} that the MFPT
 does not exist for subdiffusion. This leads us to consider
the probability distributions themselves. The method of Laplace
transforms can be used to solve for the transform of the survival
probability \cite{Gitterman1,Gitterman2} but one is left with the
very difficult task of obtaining the inverse transform, even for
the case of a sawtooth with $N=1$. However, it is shown in
appendix \ref{sec:DFPT_MFPT} that the long-time limit of the
survival probability and the first passage time distribution
scales as some power of $t$ and that they are simply related to
the expression for the MFPT
 in the context of classical diffusion as
follows
\begin{equation}
S(t) \sim \tau_{\gamma}\frac{t^{-\gamma}}{\Gamma(1-\gamma)}
\label{Soft}
\end{equation}
\begin{equation}
F(t) = -\frac{\partial S(t)}{\partial t} \sim
\tau_{\gamma}\frac{\gamma t^{-\gamma-1}}{\Gamma(1-\gamma)}
\label{Foft}
\end{equation}
Here $\tau_{\gamma}$ is the same expression as
the MFPT in classical diffusion (\ref{tauclassic0}) (or (\ref{tauclassic1})),
but containing a
generalized diffusion constant. The long-time limit is reached
when $t^\gamma \gg \tau_{\gamma} \sim L^2/D_\gamma \sim N^2$.
The relationships (\ref{Soft}) and (\ref{Foft}) ultimately arise
from the almost identical expressions for the solution $P(x,t)$ in
classical diffusion and in subdiffusion. The two solutions differ
only in the time dependence, which is an exponential for classical
diffusion.

Because the expectation time of
a first passage is infinite, any meaningful experiment, real or
computational alike, must be based on some protocol rendering
the observation time finite. We argue that in essence such a protocol is
always reduced to discarding the events which fail to come to
completion within some specified time $t^{\ast}$; in other words,
only those passage events that each complete within some time
$t^{\ast}$ are counted. The rest of the events that do not end by
time $t^{\ast}$ are terminated and discarded. The
\emph{conditional} probability distribution of the first passage
events that get counted under such a protocol is then given by
\begin{equation}
\frac{F(t)}{1-\int_{t^{\ast}}^\infty F(t)dt}
\end{equation}
For such an experiment, there exists a perfectly defined and finite
average first passage time.  This \emph{conditional} average, for
large $t^{\ast}$, is
\begin{eqnarray}
\frac{\int_0^{t^{\ast}} tF(t)dt}{1-\int_{t^{\ast}}^\infty F(t)dt}
&\sim&\frac{\tau_\gamma \frac{\gamma{t^{\ast}}^{1-\gamma}}{(1-\gamma)\Gamma(1-\gamma)}}{1-\frac{\tau_\gamma}{{t^{\ast}}^\gamma \Gamma(1-\gamma)}}\label{conditionalaverage}\\
&\sim&\tau_\gamma \frac{\gamma{t^{\ast}}^{1-\gamma}}{(1-\gamma)\Gamma(1-\gamma)}\times \nonumber\\
&&\times\left(1+\frac{\tau_\gamma}{{t^{\ast}}^\gamma
\Gamma(1-\gamma)}\right)
\end{eqnarray}
So far, the time $t^{\ast}$ should be long enough, but otherwise
arbitrary.  Now we argue that the time $t^{\ast}$ must be chosen
such that roughly about half of the passage events at a given $N$
get discarded. This requirement seems reasonable, for if one
discards a much smaller fraction $t^{\ast}$ becomes too large and
the measurements get inefficiently slow; if one discards a much
larger fraction $t^{\ast}$ becomes too small and the tail of the
distribution does not get sampled properly.  Thus, assuming half
of the events discarded, $t^{\ast}$ becomes of order
$(\tau_\gamma)^{1/\gamma}$, just at the boundary of the validity
of the asymptotics. Substituting this into
(\ref{conditionalaverage}), one obtains a scaling of
$(\tau_\gamma)^{1/\gamma} \sim N^{2/\gamma}$ for the average first
passage time.  Of course, this scaling is not unexpected for
subdiffusion with an average displacement going like $t^{\gamma/2}$.
Furthermore, due to the appearance of
the classical diffusion times $\tau_A$ and $\tau_B$
(which take the place of $\tau_\gamma$ depending on the potential)
in the average first passage time we just defined, the
asymmetry of the first passage time is once again
present in this case.

\section{Discussion}

\begin{figure}
\centerline{\scalebox{1.0}{\includegraphics{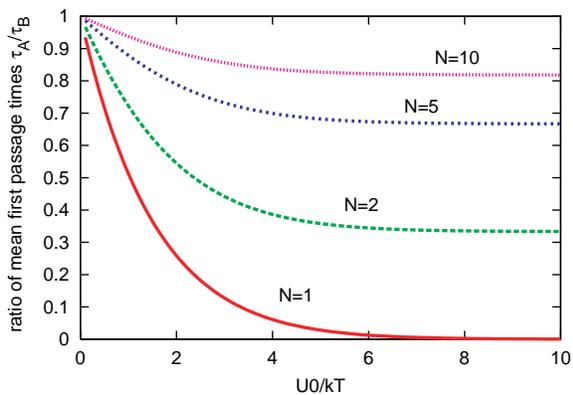}}}
\caption{(Color online) The ratio $\tau_A/\tau_B$ plotted
against the dimensionless drift or barrier height
$va/D=U_{0}/k_BT$ for $N=1$ (bottommost curve), $N=2$, $N=5$ and
$N=10$.} \label{fig:tauratioclassic}
\end{figure}

The ratio of the MFPTs in classical diffusion,
expressions (\ref{tauclassic0}) and (\ref{tauclassic1}), is plotted in figure
\ref{fig:tauratioclassic} for a few realistic values of $N$ and
$U_0$.
We see that for $U_0$ equal
to a few $k_B T$, the difference becomes small ($\sim 10 \%$) for
$N>10$. For $N=50$, corresponding to the length of ssDNA used in
the experiments by Meller and coworkers \cite{Meller_explanation},
and for $U_0 /k_B T \sim 10$, the fractional difference in MFPTs
is about $4\%$ .

We emphasize again that one cannot use the results of the
comparison between these two times (cases 1A and 1B in figure
\ref{fig:dnaporecases}) and apply it to the experimental results
in \cite{Meller_explanation} (cases 1A and 2A in figure
\ref{fig:dnaporecases}).
Due to the asymmetry of the pore, the $3^{\prime}$ and the $5^{\prime}$ threading
of DNA through one end of the pore (the so-called $cis$ side)
cannot be readily reduced to cases 1A and 1B in figure \ref{fig:sawtooth}.

Having established the difference in average first passage times
for the two asymmetric potentials, let us now turn to the scaling of the first
passage times with $N$. For large $N$, the scaling
result $N^2$ found earlier is well known for classical diffusion
or Brownian dynamics. However, this is in conflict with the
equilibration time of a polymer with $N$ monomers in the absence
of a pore and membrane, which already scales with $N$ to some
power larger than $2$ for Rouse dynamics of self-avoiding chains
\cite{KantorKardar1}.
This suggests that a correct description of
polymer translocation should be made
in the context of subdiffusion, where the scaling of the
average first passage time is to power
$1/\gamma>1$ of the classical result,
although we do not give a prediction for the value of $\gamma$
itself because the interactions involving the DNA/polymer located
outside the pore were not treated explicitly. The scaling
$N^{2/\gamma}$ is also not surprising if one takes the relation
$\left< (\Delta x)^2 \right> \sim t^\gamma$ and puts $\Delta x
\sim N$, but it does not rule out the argument made above
regarding $t^{\ast}$, only that it gives a reasonable and somewhat
expected answer. Moreover, using similar arguments,
the result $N^{2/\gamma}$ is consistent
with numerical simulations made by Chuang, Kantor and Kardar
\cite{KantorKardar1} for diffusive dynamics of self-avoiding
chains in two dimensions. They found that the average of the first
passage time scales as $\tau \sim N^{2.5}=N^{1+2\nu}$, where
$\nu=3/4$ in two dimensions. They also argued, assuming that the
translocation coordinate goes like $\left< \Delta x^2(t) \right>
\sim t^\gamma$ at short times, that $\gamma=2/(1+2\nu)$.
Eliminating $\nu$, their formulas imply that $\tau \sim
N^{2/\gamma}$.

To summarize, based on our calculations for the mean first passage
times in asymmetric sawtooth potentials and experiments by Meller
and coworkers \cite{Meller_explanation}, we expect that the
average first passage times for the four cases indicated in figure
\ref{fig:dnaporecases} are all different. The expression for the
tail of the first passage time distribution in subdiffusion is of
the form $\gamma\tau_{\gamma}/\Gamma(1-\gamma)t^{1+\gamma}$,
where $\tau_{\gamma}$ is the formula for the mean first passage
time in classical diffusion. Because the power of $t$ in the
distribution is less than two, the mean first passage time
diverges. By constructing an average from the first passage times
less than time $t^{\ast}$ such that approximately half of the
passages get rejected, we find an average that scales as
$(\tau_\gamma)^{1/\gamma} \sim N^{2/\gamma}$.

\section{Acknowledgments}

This work was inspired by an interesting seminar talk given
by Amit Meller.
We also acknowledge useful subsequent discussions with him.  RCL
acknowledges the support of a doctoral dissertation fellowship
from the University of Minnesota graduate school. We also wish to
thank the Minnesota Supercomputing Institute for the use of their
facilities.  This work was supported in part by the MRSEC Program
of the National Science Foundation under Award Number DMR-0212302.

\appendix

\section{Fractional Fokker-Planck equation}\label{sec:fractional_FFPE}

A generalization of the FPE describing anomalous diffusion is
given by the fractional FPE \cite{Metzler}
\begin{equation}
\frac{\partial P(x,t)}{\partial t}=\ _0 D^{1-\gamma}_t
L_{\mbox{FP}}P \label{FFPE1}
\end{equation}
Equivalently,
\begin{equation}
\ _0 D^{\gamma}_t P(x,t) -
\frac{t^{-\gamma}P(x,0)}{\Gamma(1-\gamma)}=L_{\mbox{FP}}P
\label{FFPE2}
\end{equation}
where the Fokker-Planck operator is defined as
\begin{equation}
L_{\mbox{FP}}=D_{\gamma}\frac{\partial}{\partial x} e^{-U(x)}
\frac{\partial}{\partial x} e^{U(x)} \label{FPOPERATOR}
\end{equation}
Here $D_\gamma$ is a generalized diffusion coefficient and $U(x)$
is an external potential. We have also set $k_BT=1$ and the
Einstein relation is implicit. The Riemann-Liouville fractional
operator is defined through
\begin{equation}
\ _0 D^{1-\gamma}_t W =
\frac{1}{\Gamma(\gamma)}\frac{\partial}{\partial t}\int_0^t
dt^{\prime}\frac{W(x,t^{\prime})}{(t-t^{\prime})^{1-\gamma}}
\label{FFPE3} \ .
\end{equation}
One can easily check that the FFPE reduces to the FPE or diffusion
equation for $\gamma=1$.

Given the initial distribution $P(x,0)=\delta(x-x_0)$, the
solution to equation (\ref{FFPE1}) is given by the bilinear
expansion \cite{Metzler}
\begin{eqnarray}
P(x,t;x_0,0) & = & e^{U(x_0)/2-U(x)/2} \times \nonumber
\\ & \times & \sum_{n=0}^{\infty} \psi_n(x) \psi_n(x_0)
E_{\gamma}(-\lambda_{n}t^\gamma)\ . \label{eigensol}
\end{eqnarray}
The functions $\phi_n(x)=e^{-U(x)/2}\psi_n(x)$ and
$T_n(t)=E_{\gamma}(-\lambda_{n}t^\gamma)$ appear in the separation
of variables ansatz $W_n(x,t)=\phi_n(x)T_n(t)$. The product
function $W_n(x,t)$ satisfies the FFPE.  Note that the coordinate
dependence comes through the eigenfunctions $\psi_n(x)$ or
$\phi_n(x)$, which are the same as for regular diffusion,
satisfying the (eigenvalue) equations
\begin{equation}
L_{\mbox{FP}}\phi_n(x)=-\lambda_{n}\phi_n(x) \ ,
\end{equation}
\begin{equation}
L^{hermitian}_{\mbox{FP}}\psi_n(x)=-\lambda_{n}\psi_n(x) \ ,
\end{equation}
\begin{equation}
L^{hermitian}_{\mbox{FP}}=e^{U(x)/2}L_{\mbox{FP}}e^{-U(x)/2} \ .
\end{equation}
However, as to the time dependence, which for classical diffusion
is described by exponentials ($e^{-\lambda_nt}$), for subdiffusion
it must satisfy the equation
\begin{equation}
\frac{d T_n(t)}{dt}=-\lambda_{n}\ _0 D^{\gamma}_t T_n(t) \ .
\label{MittagDiffEq}
\end{equation}
One can check that the following series definition of the
Mittag-Leffler function $E_\gamma(z)$ satisfies equation
(\ref{MittagDiffEq})
\begin{equation}
E_\gamma(z)=\sum_{m=0}^\infty \frac{z^m}{\Gamma(1+\gamma m)} \ .
\label{Mittagseries}
\end{equation}
This function is a natural extension of the exponential function,
to which it degenerates for $\gamma=1$.

By taking the Laplace transform of both sides of equation
(\ref{MittagDiffEq}), one obtains an alternative definition of the
Mittag-Leffler function
\begin{equation}
{\cal L}\left\{ E_{\gamma}(-\lambda t^\gamma) \right\}=\left(
s+\lambda s^{1-\gamma}\right)^{-1} \label{MittagLaplace}
\end{equation}
(The subscript in the constant $\lambda$ has been dropped.) The
long-time limit of the Mittag-Leffler function corresponds to the
small $s$ limit of the Laplace transform. Expanding
(\ref{MittagLaplace}) in a series for small $s$,
\begin{eqnarray}
{\cal L}\left\{ E_{\gamma}(-\lambda t^\gamma) \right\} &\sim& \frac{1}{\lambda s^{1-\gamma}}\left( 1-\frac{s^\gamma}{\lambda}+\left(\frac{s^\gamma}{\lambda}\right)^2 - \cdots\right) \nonumber\\
&\sim& \sum_{m=1}^{\infty} (-1)^{m+1}\frac{s^{\gamma
m-1}}{\lambda^m}
\end{eqnarray}
Taking the inverse transform, one obtains the long-time behaviour
of the Mittag-Leffler function
\begin{equation}
E_\gamma(-\lambda
t^\gamma)\sim\sum_{m=1}^\infty\frac{(-1)^{m+1}}{\Gamma(1-\gamma
m)}\left(\lambda t^\gamma \right)^{-m} \label{mittaglimit}
\end{equation}

For $\epsilon=1-\gamma=0$, the Laplace transform
(\ref{MittagLaplace}) becomes $(s+\lambda)^{-1}$, the inverse
transform of which is an exponential. For $\epsilon$ close to $0$,
we expect a long time interval in which the behaviour of the
Mittag-Leffler function $E_{\gamma}(-\lambda t^\gamma)$ behaves
like an exponential; at much longer times the behaviour changes to
a power law. The crossover is expected to happen when $e^{-\lambda
t_c} \sim \frac{1}{\Gamma(\epsilon)\lambda t_c^\gamma}$, or at
about $t_c \sim \frac{1}{\lambda}\ln(\frac{1}{\epsilon})$.

\section{Differential equation satisfied by the mean first passage
time}\label{sec:equation_for_tau}

Recall that the MFPT can be calculated from (equation
(\ref{MFPT}))
\begin{equation}
\tau(x_0)=\int_0^{\infty}\int_{-L}^{L}P(x,t)dxdt \label{MFPT2}
\end{equation}

To derive an ordinary differential equation satisfied by
$\tau(x_0)$, apply the operator
$e^{U(x_0)}L_{\mbox{FP},x_0}e^{-U(x_0)}$ to equation (\ref{MFPT2})
and use the eigenfunction expansion solution (\ref{eigensol}) for
$P(x,t)$,
\begin{widetext}
\begin{eqnarray}
e^{U(x_0)}L_{\mbox{FP},x_0}e^{-U(x_0)}\tau(x_0) & = & \int_0^\infty\int_{-L}^L e^{U(x_0)/2-U(x)/2}\sum_{n=0}^{\infty}(-\lambda_{n})\psi_n(x)\psi_n(x_0)E_{\gamma}(-\lambda_{n}t^\gamma) dxdt\nonumber\\
& = & \int_0^\infty\int_{-L}^L L_{\mbox{FP}}P(x,t)dxdt =
\int_0^\infty\int_{-L}^L \left[ \ _0 D^{\gamma}_t P(x,t) -
\frac{t^{-\gamma}P(x,0)}{\Gamma(1-\gamma)} \right] dxdt\nonumber
\end{eqnarray}
\end{widetext}
In the last two steps, the eigenvalue equations and the second
version of the FFPE (equation (\ref{FFPE2})) was used.

Using the initial condition $P(x,0)=\delta(x-x_0)$ and the
definition of the fractional operator, after some algebra one
obtains
\begin{eqnarray}
&&e^{U(x_0)}L_{\mbox{FP},x_0}e^{-U(x_0)}\tau(x_0)=\nonumber\\
&&-\lim_{t\rightarrow \infty} \left[
\frac{t^{1-\gamma}}{\Gamma(2-\gamma)} -
\frac{1}{\Gamma(1-\gamma)}\int_0^t
\frac{S(t^{\prime})}{(t-t^{\prime})^\gamma}dt^{\prime} \right]
\end{eqnarray}
or
\begin{eqnarray}
&&D_{\gamma}e^{U(x_0)}\frac{\partial}{\partial x_0} e^{-U(x_0)} \frac{\partial}{\partial x_0}\tau(x_0)=\nonumber\\
&&-\lim_{t\rightarrow \infty} \left[
\frac{t^{1-\gamma}}{\Gamma(2-\gamma)} -
\frac{1}{\Gamma(1-\gamma)}\int_0^t
\frac{S(t^{\prime})}{(t-t^{\prime})^\gamma}dt^{\prime} \right]
\end{eqnarray}
For $\gamma=1$, corresponding to classical diffusion, the survival
probability $S(t^{\prime})$ decays exponentially and the term with
the integral goes to zero, yielding the familiar result of $-1$
for the right-hand-side
\cite{Pontryagin,FiftyYearsKramers,Risken}. For $\gamma<1$,
$S(t^{\prime})$ goes like $(t^{\prime})^{-\gamma}$ (see
(\ref{S-series})) and the term with the integral goes like
$t^{1-2\gamma}$. The right-hand-side diverges, which hints at the
non-existence of the MFPT for subdiffusion
\cite{Yuste,Gitterman2,WhenAnomalous,Restaurant,PhysicsWorld}.

\section{Solution for the mean first passage time in a sawtooth potential}\label{sec:solution}
From the previous section, the differential equation satisfied by
the MFPT in the context of classical diffusion is (temporarily
putting back $k_BT$)
\begin{equation}
De^{U(x)/k_B T}\frac{d}{dx}e^{-U(x)/k_B T}\frac{d\tau(x)}{dx}=-1
\end{equation}
We solve for $\tau(x)$ in this equation for a sawtooth potential
(case A, figure \ref{fig:sawtooth}) subject to the boundary
conditions $d\tau/dx(0)=0$ and $\tau(L)=0$, and the continuity of
$\tau$ and $e^{-U/k_B T}d\tau/dx$ in $(0,L)$. In what follows we
let $\xi=e^{va/D}=e^{U_{0}/k_B T}$.

The solution, for $x$ between $(m-1)a$ and $ma$ where $m$ is an
integer between $1$ and $N$ (inclusive), is given by
\begin{equation}
\tau(x)=A_{m}e^{-vx/D}-\frac{x}{v}+B_m
\end{equation}
The coefficients $A$ and $B$ are given by
\begin{eqnarray}
A_m & = & \xi^{m-1}\frac{D}{v^2}\left[(\xi-1)(m-1)-1\right] \ ;
\nonumber \\
B_m & = &
\frac{D}{v^2}+N\left(\frac{a}{v}-(1-1/\xi)\frac{D}{v^2}\right)+
\nonumber \\ & + &
\frac{D}{v^2}\frac{(\xi-1)^2}{\xi}\frac{(N-m)(N+m-1)}{2} \ .
\end{eqnarray}
The MFPT for a particle initially located at $x=0$ is given by
$\tau(0)=A_1+B_1$. To obtain the solution for case B in figure
\ref{fig:sawtooth}, we may flip $\xi$ ($\rightarrow 1/\xi$) and
the sign of $v$ in the expressions above.

\section{Relationship between the DFPT in anomalous diffusion and the MFPT in classical diffusion}\label{sec:DFPT_MFPT}
Since the MFPT does not exist for subdiffusion, one would want to
calculate the distributions instead. In the long time limit, using
(\ref{eigensol}) and (\ref{mittaglimit}),
\begin{widetext}
\begin{equation}
\lim_{t\rightarrow \infty} P_{\gamma}(x,t;x_0,0)\sim \sum_{n}
e^{U(x_0)/2-U(x)/2}\psi_n(x)\psi_n(x_0)\frac{1}{\Gamma(1-\gamma)\lambda_{n}t^\gamma}
\end{equation}
\begin{equation}
\lim_{t\rightarrow \infty} S_{\gamma}(t)\sim \sum_{n}
\left(\int_{-L}^L e^{-U(x)/2}\psi_n(x)dx \right)
e^{U(x_0)/2}\psi_n(x_0)\frac{1}{\Gamma(1-\gamma)\lambda_{n}t^\gamma}
\label{S-series}
\end{equation}
\begin{equation}
\lim_{t\rightarrow \infty} F_{\gamma}(t)\sim \sum_{n}
\left(\int_{-L}^L e^{-U(x)/2}\psi_n(x)dx \right)
e^{U(x_0)/2}\psi_n(x_0)\frac{\gamma}{\Gamma(1-\gamma)\lambda_{n}t^{\gamma+1}}
\label{F-series}
\end{equation}
\end{widetext}
Again, these results indicate that the MFPT diverges for
$\gamma<1$. It is also interesting to note that all the
eigenfunctions $\psi_n(x)$, not just the ground state, enter in
the expressions.

To make sense of the expression multiplying
$\frac{\gamma}{\Gamma(1-\gamma)t^{\gamma+1}}$ in (\ref{F-series})
write down the corresponding solution for classical diffusion
under the same potential and the same value for the diffusion
coefficient
\begin{eqnarray}
P(x,t;x_0,0) & = & e^{U(x_0)/2-U(x)/2} \times \nonumber
\\ & \times & \sum_{n=0}^{\infty}\psi_n(x)\psi_n(x_0)\exp(-\lambda_{n}t)
\end{eqnarray}
(Note exponential instead of Mittag-Leffler function). The
survival probability is given by
\begin{eqnarray}
S(t) & = & \sum_{n} \left(\int_{-L}^{L} e^{-U(x)/2}\psi_n(x)dx
\right) \times \nonumber \\ & \times &
e^{U(x_0)/2}\psi_n(x_0)\exp(-\lambda_{n}t)
\end{eqnarray}
While the MFPT is given by
\begin{eqnarray}
\tau(x_0)&=&\int_0^\infty S(t)dt\nonumber \\
& = & \sum_{n} \left(\int_{-L}^{L} e^{-U(x)/2}\psi_n(x)dx \right)
\times \nonumber \\ & \times & e^{U(x_0)/2}\psi_n(x_0)\frac{1
}{\lambda_{n}}
\end{eqnarray}
which is identical to the coefficient of $\gamma \left/
\Gamma(1-\gamma)t^{\gamma+1} \right.$ in (\ref{F-series}).

\section{Effective diffusion constant in the steady state}\label{sec:steadystate}
In this section we determine the steady state current $J$ given
fixed concentrations $c(0)$ and $c(L)$ at the boundaries. Let the
potential $U(x)$ satisfy $U(0)=U(L)=0$, but is otherwise
arbitrary. The classical diffusion equation is given by
\begin{equation}
\frac{\partial c}{\partial t}=-\frac{\partial J}{\partial x}
\end{equation}
where $J=-D e^{-U(x)} \frac{\partial}{\partial x} e^{U(x)}c$ (see
equation (\ref{FPE})). In the steady state, $\frac{\partial
c}{\partial t}=0$, which implies that $J$ is spatially uniform.
Integrating $J e^{U(x)}=-D\frac{\partial}{\partial x}
e^{U(x)}c(x)$ and utilizing the boundary conditions, one obtains
\begin{equation}
J=\frac{D}{\int_0^L e^{U(x)}dx}(c(0)-c(L)) \ .
\end{equation}
This expression is identical with Fick's law with an effective
diffusion constant of $\frac{DL}{\int_0^L e^{U(x)}dx}$.

\end{document}